# Quantum control in artificial neurons with superconductor-ionic memory inserted in the feedback

**Authors:** Osama M. Nayfeh[1*], Haik Manukian[1], Matthew Kelly[1], Justin Mauger[1]

**Affiliations:**
  [1] US Naval Information Warfare Center Pacific (NIWC PAC), San Diego, CA

**\*** Corresponding Author: osama.m.nayfeh.civ@us.navy.mil

**Abstract:** To improve artificial intelligence/autonomous systems and help with treating neurological conditions, there's a requirement for artificial neuron hardware that mimics biological. We examine experimental artificial neurons with quantum tunneling memory using 4.2 nm of ionic Hafnium oxide and Niobium metal inserted in the positive and negative feedback of an oscillator. These neurons have adaptive spiking behavior and hybrid non-chaotic/chaotic modes. When networked, they output with strong itinerancy. The superconducting state at 8.1 Kelvin results in Josephson tunneling with signs that the ionic states are influenced by quantum coherent control in accordance with quantum master equation calculations of the expectation values and correlation functions with a calibrated time dependent Hamiltonian. We experimentally demonstrate a learning network of 4 artificial neurons, and the modulation of signals.

**One-Sentence Summary:** Adaptive behavior, itinerancy, and quantum control in an artificial neuron with two superconductor-ionic tunneling memories.

**Main Text:**

## Introduction

Harnessing and merging the fields of neuromorphic computing and quantum information processing is a practical way to take steps towards designing and developing new hardware that operate with biological inspired autonomous control *(1)*. Therefore, increasing our understanding and knowledge of neuron behavior as well as developing technologies that can mimic it, are essential to move forward with this goal *(2)*. The current approaches typically emulate neuron models with digital and analog transistor circuits *(3)* and have achieved impressive results, however, they may lack sufficient built-in memory, non-linear dynamical behavior, and quantum phenomenon in their operation to come close to biological functionality. Thus, integrating memristors and other novel memory devices in the design of artificial neuron hardware is an intense topic in the scientific community *(4)*. Recently, memristive circuits were demonstrated that output various spiking modes, but observation of hybrid chaotic modes and itinerancy require increased non-linearity and feedback *(5)*. Theoretical proposals have also recently pointed to the possibility of harnessing quantum mechanical phenomenon in the design of artificial neurons to take advantage of quantum information science protocols, but experimental demonstrations are limited *(6-8)*.

We designed and experimentally examined a high-speed artificial neuron circuit and simple networks with quantum tunneling memory hardware inserted in the positive and negative feedback loops of an analog spiking oscillator, providing inhibitory and excitatory dynamics in a coupled non-linear dynamical system with built-in synaptic memory. We demonstrate the adaptive properties, hybrid non-chaotic/chaotic modes and the emergence of strong itinerant behavior when networked. With the memories in the superconducting state, Josephson tunneling emerges and results in a new solid-state superconductor-ionic system where signs of the influence of coherent quantum control are observed in accordance with quantum master equation calculations of the expectation values and correlation functions using a calibrated time dependent Hamiltonian under strong driving field conditions. As a utility of this hardware, we experimentally demonstrate a 4-neuron ring network performing autonomous learning and the modulation of signals that exploit the biologically inspired functionality.

## Design of artificial neurons

The schematic of the artificial neuron circuit we examined is shown in Fig. 1(**A**)**.** We inserted Nb-HfO$_x$-Nb quantum tunneling memory elements into the positive and negative feedback loops of an analog Wien/relaxation oscillator. At room temperature, these memories operate as memristive elements due to the ionically active HfO$_x$ *(9)* and when cryogenically cooled into the superconducting state, operate as a superconductor-ionic quantum memory that also use Josephson tunneling. During the dynamical operation of this circuit, the memory states in the feedback loops are adjusted by the bias voltages and level of current flow. The resistances and capacitances help adjust the spiking rates, the operational amplifier enables feedback and coupling between the memristors, and the diode provides stabilization. Inhibitory and excitatory impacts to the memory states and their non-linearity enable adaptive spiking as the output response. To obtain insights on the operation, we first examined the room/warm temperature response and derived with a Kirchoff's voltage and current law nodal analysis *(see supplementary)* a full analytical model for the output $v_o(t)$:

$$\frac{d^2 v_o(t)}{dt^2} + \left(\frac{R_{Mem2}(t)}{R_3} + \frac{C_1}{C_2} + \frac{dR_{Mem2}(t)}{dt}C_1 + \frac{R_{Mem1}(t)}{R_4}\right)\frac{1}{R_{Mem2}(t)C_1}\frac{dv_o(t)}{dt} + \cdots$$
$$\cdots\left(\frac{1}{R_3 C_2} + \frac{1}{R_3}\frac{dR_{Mem2}(t)}{dt}\right)\frac{1}{R_{Mem2}(t)C_1}v_o(t) = 0 \tag{1}$$

and this non-linear differential equation is solved numerically while coupled with the equations for the memory dynamical state variables $x_{Mem1}(t)$ and $x_{Mem2}(t)$ and the mem-resistances $R_{Mem1}(t)$ and $R_{Mem2}(t)$ using the model in *(10)* as a starting point:

$$\frac{dx_{Mem}(t)}{dt} = \lambda[\eta_1 \sinh(\eta_2 V(t)) - \frac{f(V,x,T)x_{Mem}(t)}{\tau}] \tag{2}$$

$$R_{Mem}(t) = V(t)/\left(\left(1 - x_{Mem1}(t)\right)\alpha[1 - e^{\beta V(t)}] + x_{Mem1}(t)\gamma \sinh(\delta V(t))\right) \tag{3}$$

Where the hyperbolic dependencies capture the voltage *V(t)* dependent ionic drift-diffusion processes in metal-oxide memristors. The parameters $\gamma$, $\delta$, $\alpha$, $\beta$, are pre-factors and exponents that adjust the levels of quantum tunneling and Schottky emission and $\lambda$, $\eta_1$, $\eta_2$ are adjustable state variable pre-factors. $\tau$ is a diffusion constant that represents the level of short- and long-term memory i.e., non-volatility *(10)*. We introduced a functional dependence $f(V, x, T)$ to capture the non-linear dependence of the ion velocity with temperature *(11)* in accordance with the expression $v \approx ae^{-\frac{Ua}{k_bT}}sinh\frac{qEa}{2k_bT}$, where $U$ is an activation energy, $a$ the ion periodicity, $E$ is the electric field strength and $k_bT$, the thermal energy. The resistances and capacitances of the circuit were adjusted to take into the account the memory elements inserted and $C_{1-2}$, and $R_{1-4}$ were selected within the 100 nF-1 μF and 1-10 kOhm ranges to obtain a MHz speed response. Fig. 1(**B**) shows the output of this model, with a constant $\tau$ that produces classical spiking and Fig. 1(**C**) with a strong non-linear functionality that results in a hybrid chaotic mode. Fig. 1(**D**) shows the phase-plane trajectories where signs of itinerancy are apparent.

To provide further insights into the learning operations, larger scale circuit simulations (see methods) of a network with 9 artificial neurons as shown in Fig. 1(**E**) were also done with the model parameters in Table S1. Fig. S2 shows several examples of various spiking modes that can be obtained for a single artificial neuron. Fig. 1(**F,G**) shows the simulated output and the dynamics of the memristor dynamical state variables upon the application of a small sinusoidal stimulus signal. The output starts with adaptive training and finally to a steady state spiking output that represents the learned encoded information that was reinforced by the autonomous learning as determined from the memristors dynamical state variables. We note that such simulations are limited to a scale of around 40 artificial neurons on a digital computer, thus further motivating the development of dedicated AI hardware.

### Adaptive and hybrid behavior

We experimentally examined this artificial neuron circuit using memristor/resistive switching (ReRAM) hardware we built on silicon wafers with 4.2 nm of atomic layer deposited (ALD) ionically active Hafnium-oxide (HfO$_x$), and sputtered Niobium (Nb) electrodes i.e. (Nb-HfO$_x$-Nb) produced with the process in *(9)*. First, memristors were independently characterized. Fig. 2(**A**) and Fig. S3 show pulsed current/voltage (IV) measurements taken at room temperature sweeping forward and reverse from -2.0 to 2.0 V with 300/900 ns pulse widths and 60/100 ns rise/fall times. The characteristics are non-linear with a significant hysteresis and resistance switching of current of 4-5 orders of magnitude. The primary conduction mechanism for this voltage range and with 4.2 nm of HfO$_x$ and a Nb-HfO$_x$ barrier of 2.2 eV is by direct and field-effect quantum tunneling as determined from the dependency from a plot of $ln(I/V^2)$ vs. $1/V$ as shown in Fig. 2(**B**) where regimes of direct tunneling (DT) with a transition to field-effect tunneling is evident *(9,12)* as shown in Fig. 2(**C**) presents a representative band diagram in accordance with the expected tunneling electron flow between normal metals at these warm temperatures, where $T_{Nb-HfOx}$ is the transmission probability that can be understood from a Schrödinger solution and we consider the presence of the ionic states impacting it and hence, $I = \frac{2\pi A}{\hbar}\int_{-\infty}^{+\infty}|T_{Nb-Hfox}|^2 N_{Nb}(\varepsilon - V)N_{Nb}(\varepsilon)[f(\varepsilon - V) - f(\varepsilon)]d\varepsilon$ *(13)* where N$_{Nb}$ are the density of electron states available for tunneling and $f$ is the Fermi-Dirac function. The changes in the memory response with increasing pulsing speed, point to a combination of effects. The hysteresis is consistent with volatile ionic changes/motion, due to the creation and motion of ionic oxygen vacancy states and any residual thermal fluctuations *(14-17)*. The increasingly larger hysteresis of 0.5-0.7 V with 300 ns and 60 ns pulsing and the more pronounced shoulder is indicative of negative differential resistance and the combined contributions of the partial ferro-electric effects and metal-insulator transitions due to the ionic composition and the Nb-HfO$_x$ interfaces *(18-19)*.

We measured the full circuit response with the memristors inserted. The experiments were done with the memristors probed on chip while in a cryogenic probe station and wired up with discrete op-amps, resistors, capacitors, and diodes that resided on an electronics breadboard, Fig. S4. Fig. 2(**D**) shows example experimental outputs with changing DC bias voltages. First, we isolated a spiking mode that is classical and stable in appearance to those produced with the Hodgkin-Huxley (HH) neuron model *(20,21)*. Next, we adjust the DC stimulus to change the response to a chaotic mode and a bursting mode. The extracted spiking rates for these modes is greater than 200 kHz with MHz spectrum. To examine the degree of spike rate adaption (SRA) and other adaptive behaviors which are common attributes in biological neurons, and key for training and learning, we also examined the addition of a small signal excitation. Fig. 2(**E**), compares the outputs with just DC voltage stimulus and with an added RF sinusoidal stimulus (0.2 V$_{pp}$/5 MHz) and Fig. 2(**F**), the extracted spiking rates versus time. This rate adapts starting from around 200.1 kHz and reduces over the course of around half a milli-second to 100.3 kHz with a dependence like that produced with the biologically inspired adaptive exponential model (AdEx) *(22)*. With the added RF analog stimulus, the spiking rates start at 201.3 kHz and reduce to 160.2 kHz with a more abrupt transition that effectively produced a new bursting spiking mode as the adaptive response. An example of the phase-plane trajectories of a single artificial neuron at room temperature that produces hybrid non-chaotic/chaotic attractor modes is shown in Fig 2(**G**) and Video S1. With a single neuron, the attractors enter hybrid non-chaotic/chaotic modes with a distinct saddle point *(23,24)* and as we add neurons the complexity observably increases consistent with the formation of a neural network as shown in the examples shown in Fig. 2(**H-K**) and Videos S2-S4.

## Quantum control

Next, we examined experimentally the effect of cryogenically cooling the memories into the superconducting state. Fig. 3(**A**) show measured data of a Nb-HfO$_x$-Nb device taken at 8.1 Kelvin sweeping from +/- 1.0 mV in the forward and reverse directions. A critical current due to cooper pair tunneling, a sum-gap voltage $\Delta$=0.31 mV and at increasing voltage a quasiparticle current. These characteristics have a hysteretic memory effect. With 0.24 Tesla applied, a decreased tunneling current is apparent and consistent with a Josephson tunneling junction with Nb electrodes *(12)*. The characteristics are impacted by the excitation and decay of carriers into ionic centers where there is energy level splitting in the quantized environment. Fluctuations are attributed to residual noise with a thermal energy of 0.6 meV. The impacts of the cryogenic temperature and field effect further impacts the coherence length $\xi_n = \left(\frac{D}{k_B T}\frac{\hbar}{2\pi}\right)^{1/2}$ through the diffusion constant $D=(1/3)v_f l_n$ with Fermi velocity $v_f$, $l_n$ the mean free path depending on the level of ionic states *(13),* and $\hbar$ the reduced Planck's constant resulting in observable hysteresis.

We then examined the output of the artificial neuron circuit, while the memories were in this superconducting state. We used periodic pulsed excitation with 0.4-0.5 milli-second widths and Fig. 3(**B**, **C**) presents two examples of the collected experimental spectra where we plot the output voltages at the two respective memory nodes vs. time in Fig. 3(**D**) and Video S5 are raw oscilloscope captures and, Fig.3 (**E**, **F**) in the phase-space (Video S6) for increasing excitation strength and when a network is formed. The dynamical behavior observed in the experimental spectra appeared distinctly different and influenced from quantum phenomenon occurring in the superconducting state and outside what can be modeled with equations 1-3 for the warm temperature situation. To investigate further this notion, we performed calculations with the quantum master equation (QME) *(25)* where we model this solid-state system as a quantum one where the driving field is affected by the Josephson tunneling, that impacts the energy flow between the ionic states in the HfO$_x$. We created a modified type of Hamiltonian to represent transitions between ground $|g>$, excited $|e>$ states and their respective interactions with the driving field. We introduced a strong-field time dependent function with form $\sin\left(\left(\frac{t}{e}\right)^2\right)$ to approximate the spiking oscillator backbone and thus:

$$H_{super-ionic}(t) = -g\left(\sigma_{ge}^{\dagger}a + a^{\dagger}\sigma_{ge}\right) - A(\sigma_{ge}^{\dagger} + \sigma_{ge})\sin\left(\left(\frac{t}{e}\right)^2\right) \tag{4}$$

where *g* is the coupling strength, *A* the intensity and *e* the frequency of the driving field and these parameters. As a reminder, the QME describes the evolution of the density matrix

$$\rho(t): \frac{d\rho(t)}{dt} = -\frac{i}{\hbar}\left[H_{super-ionic}(t), \rho(t)\right] + \sum_n \frac{1}{2}\left[2C_n\rho(t)C_n^{\dagger} - \rho(t)C_n^{\dagger}C_n - C_n^{\dagger}C_n\rho(t)\right] \tag{5}$$

Dissipation is introduced through the collapse operators $C_n$ and $C_n^{\dagger}$ and we adjusted the rate of decay to 0.15 and simulated two conditions (I) where the intensity of the driving field *A*=10 and a Hilbert space dimension of 4 and (II) A=40 with a Hilbert space of 8 and plot the time evolution of $<a^{\dagger}a>$ and $<\sigma_{ge}^{\dagger}\sigma_{ge}>$ and also plotted against each other to represent the experimental data collected in the phase-space. Fig. 3(**G,H**) shows the calculations for scenario *(I)* and Fig. 3(**I,J**) for scenario (II)**.** The behavior in these simulations is in good agreement with that observed experimentally. The larger Hilbert space required to capture the effects of the increasing driving field in the experiments points supports the notion that quantized energy level splitting is further enhanced and thus increases the dimensionality of the ionic states available for any coherent processes. To further evaluate, we calculated the first and second order correlation functions $g^{(1)}(t)$ and $g^{(2)}(t)$ as implemented in *(25)*, that represent the level of quantum coherence and entanglement that could be supported: $g^{(1)}(t) = \frac{\langle a^{\dagger}(t)a(0)\rangle}{\sqrt{\langle a^{\dagger}(t)a(t)\rangle\langle a^{\dagger}(0)a(0)\rangle}}$ and $g^{(2)}(t) = \frac{\langle a^{\dagger}(0)a^{\dagger}(t)a(t)a(0)\rangle}{\langle a^{\dagger}(0)a(0)\rangle^2}$. The calculated correlation functions are shown in Fig. 3(**K,L**) and their sustainment over the course of duration provides further evidence that quantum coherent control occurs under these conditions. We note that while this Lindblad formulation of the QME in equation 5 is appropriate for short-term memory effects i.e., Markovian our collected experimental data show some signs under network conditions *(26,27)* that point to a time-delay reversal phenomenon followed by re-establishment that occurs in the dynamical behavior as shown by the time evolution in Fig. 3(**F**).

**Strong itinerancy and networks/applications**

Finally, we created a neuron-neuron network where the memories of one artificial neuron are kept warm at room temperature and the second neuron with its memories operated in the quantum control regime as shown in Fig. 4(**A**) and we optimized the circuit for MHz speeds. The goal was to provide a system where strong itinerant behavior becomes significantly pronounced by networking neurons with very different intrinsic attractor modes and regimes of bifurcation, as recently proposed in theoretical studies on designing artificial autonomous systems with chaotic itinerancy *(24, 28)*. Fig. 4(**B**) shows an example spiking output and phase portraits when the memristors are cooled to 8.1 Kelvin. As shown in Fig. 4(**C-E**), as temperature is reduced, the two neuron networks collective behavior enters several regimes where the attractors show strong signs of itinerancy i.e., behavior where time is spent in the trajectory of a dynamical attractor mode and then changes to several other modes (Videos S7-S9). Such behaviors are commonly seen in the collective effects of biological neurons *(29-30)*, and where for example such influence can result in spontaneous changes in motion and directionality of birds*.* Collected time -domain spectrum is shown in Fig. 4(**F**) and three-dimensional plots in Fig. 4(**G, H**).

As an application, we experimentally examined a four synthetic neuron ring feedback network proposed by the neuroscience community *(31)* as a possible way that biological organisms contribute to implementing gait control or central pattern operations as connected as shown in the photo of the breadboard Fig. S5 in a ring architecture and we provide stimulus signals as a means of introducing data for the circuit and collected the outputs at all 4 nodes in real-time Fig. 5(**A**). We isolated as an example a mode and as shown in Fig. 5(**B**) and (Video S10) and under close examination there are interesting changes to the level of phase shift happening as a function of time between the various nodes, which is the typical requirement for such a circuit to train and learn various gait or central control patterns consistent with the simulations done earlier in Fig. 1(**F, G**). With suitable driving of the circuit, we observe adaptive changes to the spiking modes because of the adaptive and itinerant properties Fig. 5(**C**) even at the scale of networks with just 4 artificial neurons. As an additional near-term utility, we use the neuron output to modulate RF Photonics communications signals. The output of an artificial neuron Fig. 5(**D**) serves as the source of I/Q modulation of a higher frequency carrier signal at 20 MHz produced with a vector signal generator (VSG). This neuron modulated RF signal then drives an acousto-optic modulator (AOM) that modulates a near infrared 975 nm fiber optic laser operating with a power of 40 mW. The modulated optical signal is then propagated in a single mode fiber and is detected by an avalanche photodetector and the output is viewed on an oscilloscope**.** As observed, this protocol results successfully in a modulated output with a complex time-dependent pattern due to the hybrid non-chaotic/chaotic initial spiking mode with an increased spiking rate of 6-10 times, Fig. 5(**E**). Therefore, such a protocol using these synthetic neurons can be used to generate complex types of signal modulations.

## Conclusions

An artificial neuron was designed and examined experimentally with quantum tunneling memory hardware formed from 4.2 nm of atomic layer deposited ionic Hafnium oxide and Niobium metal and inserted in the positive and negative feedback of an analog spiking oscillator. When operated at room/warm temperatures these memories have memristive properties and enabled the artificial neuron circuits to produce adaptive spiking behavior in accordance with biological models and a non-linear dynamical model we derived that supports the contributions of the inhibitory and excitatory feedback and the memory state non-volatilities.  When networks were formed, pronounced hybrid chaotic/non-chaotic modes with increased complexity are observed and itinerant behavior emerged. Measurements of these artificial neurons when the Nb-HfO$_x$-Nb memories are cryogenically cooled into the superconducting Josephson tunneling regime at 8.1 Kelvin revealed the influence of quantum coherent control in this solid-state superconductor-ionic system and quantum mechanical calculations of the expectation values and correlation functions with the introduction of a strong time dependent Hamiltonian are in good agreement with the acquired experimental spectrum. We demonstrated a 4-artificial neuron feedback ring network performing autonomous learning and the use of these artificial neurons for the modulation of RF photonics communications signals.

**Acknowledgments:** The use of the UCSD Qualcomm Nano³ facility for nanofabrication is acknowledged. We are grateful for the NIWC information technology and lab infrastructure. Thanks to E. Bozeman for providing some of the discrete electronics. This manuscript is a tribute to the memory of Prof. Judy L. Hoyt (MIT).

**Funding:** This research was funded by the Office of the Secretary of Defense (OSD) applied research for the advancement of priorities (ARAP) programs on quantum science and engineering (QSEP) and neuromorphic electronics (Neuro-pipe), the NIWC PAC Naval innovative science and engineering program (NISE) and the Office of Naval Research (ONR) Independent laboratory initiative for research program (ILIR). Distribution Statement A. Approved for public release: distribution is unlimited.

**Author contributions:** Conceptualization: OMN, Methodology: OMN, Investigation: OMN, HM, MK, JM Validation: OMN, Formal Analysis: OMN, Resources: OMN, Data Curation: OMN, Visualization: OMN, Writing – original draft: OMN, Writing – review & editing: OMN, HM, MK, JM

**Competing interests:** The quantum memory technology is protected by US Patents (1) Advanced process flow for quantum memory devices and Josephson junctions with heterogeneous integration", US Patent 9,455,391 and (2) Quantum memory device and method, US Patent 9,385,293.

**Data and materials availability:** All data are available in the main text or the supplementary materials." Several videos are included in the auxiliary supplementary materials. Requests for additional data or questions should be made to the corresponding author.

## Supplementary Materials

Materials and Methods

Supplementary Text

Figs. S1-S5

Videos S1 to S10

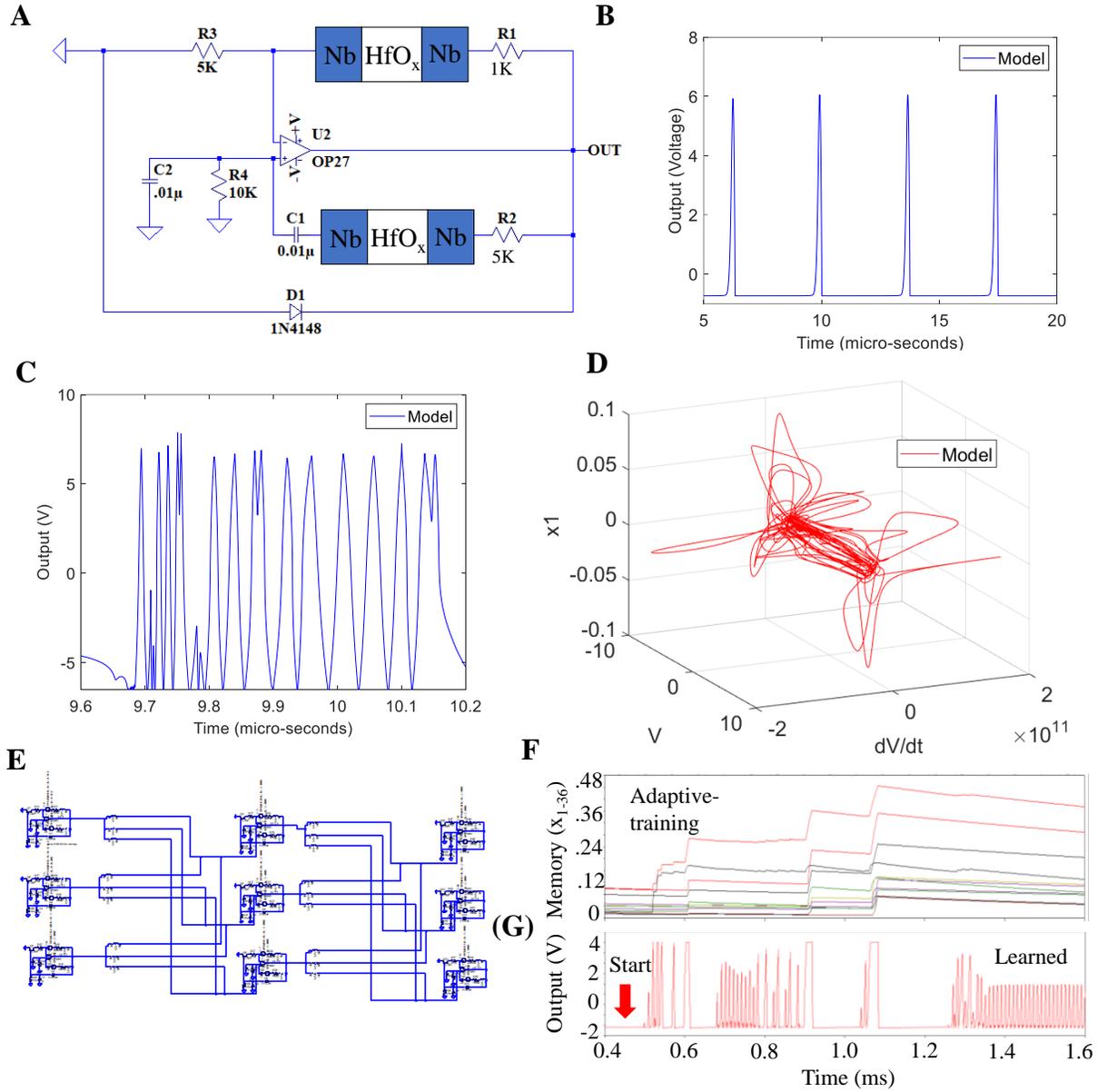

**Fig. 1. Design of artificial neurons.** (**A**) Schematic of artificial neuron circuit with Nb-HfO$_x$-Nb memories inserted in the positive and negative feedback loops along with resistors R1-R4, capacitors C1-C2, diode and operational amplifier. (**B**) Modeled output with a stable classical spiking response (**C,D**) Modeled output with strong non-linearity and in the phase-plane with signs of hybrid non-chaotic/chaotic behavior and itinerancy (**E**) Schematic of a simulated network of 9 of these artificial neurons and 36 memristors (**F**) The dynamics of all the inserted memories during training and learning operations and (**G**) The spiking output response after

introduction of a small signal stimulus signal demonstrating the ability to adapt the output during training and the final learned output.

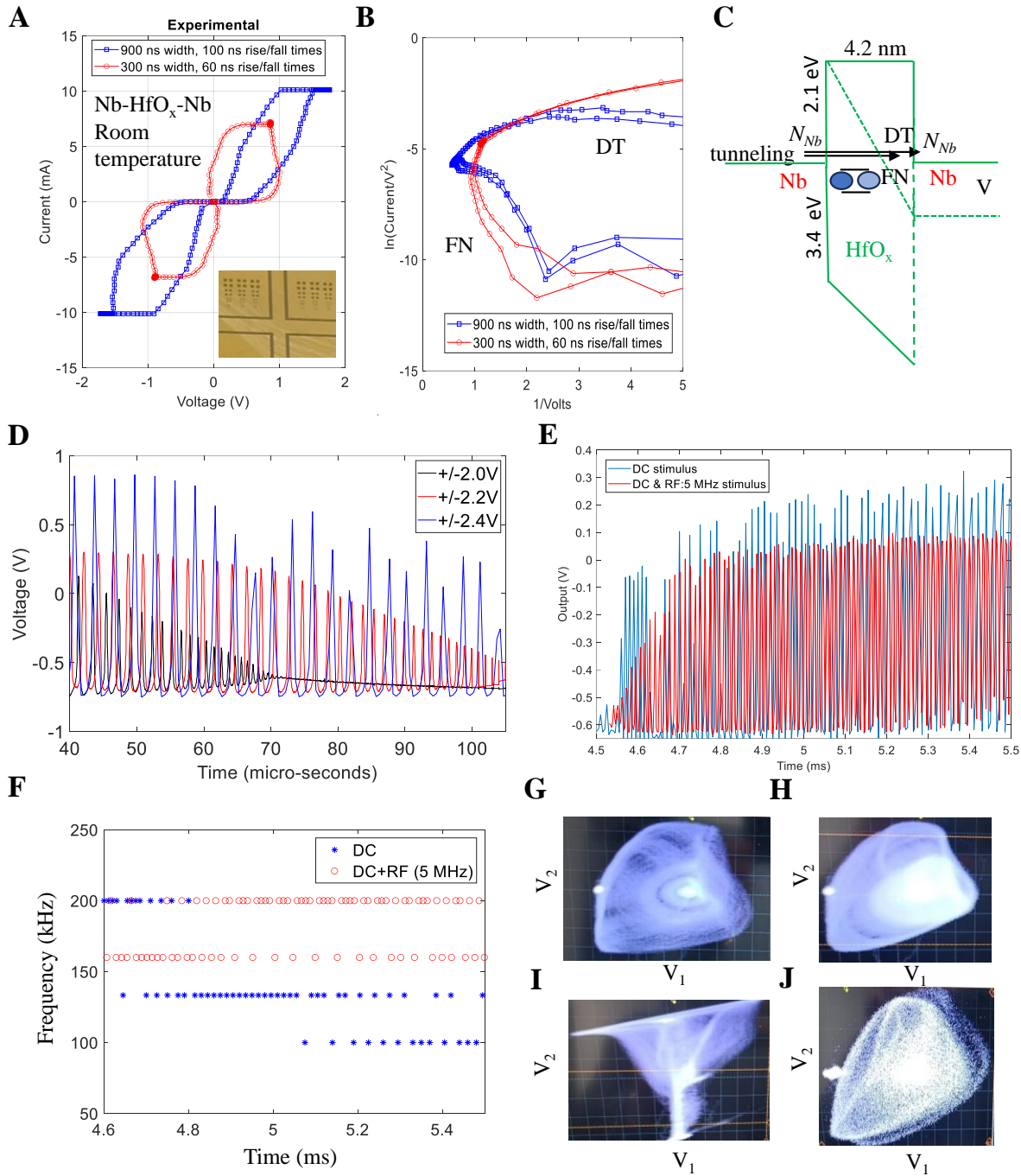

**Fig. 2. Adaptive and hybrid spiking behavior.** (**A**) Pulsed I/V measurements of Nb-HfO$_x$-Nb memory at room temperature with memristive behavior (**B, C**) Extraction of direct and field-effect tunneling contributions in accordance with energy-band diagram (**D**) Measurements of artificial neuron circuit with varying the DC and (**E**) with an added 5 MHz RF stimulus signal (**F**) Extraction of the spike rate adaptation for DC and DC+RF stimulus cases (**G**) Oscilloscope

capture in phase space with DC+RF stimulus (**H-J**) Oscilloscope capture examples in phase-plane from one artificial neuron to two with the formation of a network.

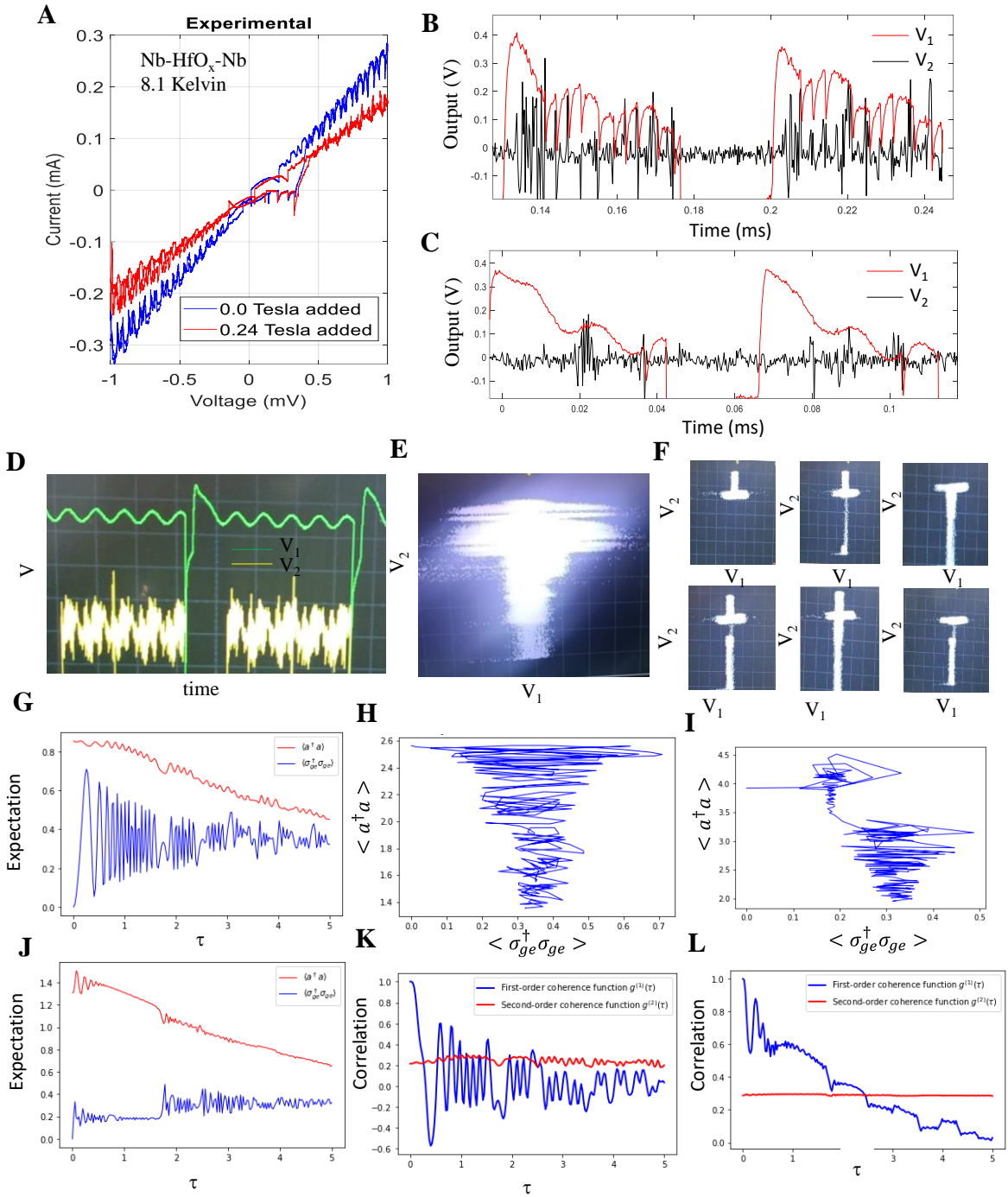

**Fig. 3. Quantum control.** (**A**) Measurements of Nb-HfO$_x$-Nb memory at 8.1 Kelvin (**B, C**) Artificial neuron circuit output with memories in the superconducting state with pulsed excitation (**D**) Oscilloscope capture in the time-domain and (**E, F**) Phase-plane (**G, H**) Modeled expectation values with the quantum master equation with a Hilbert space of 4 and (**I, J**) 8 and (**K, L**) The extracted first and second order correlation functions.

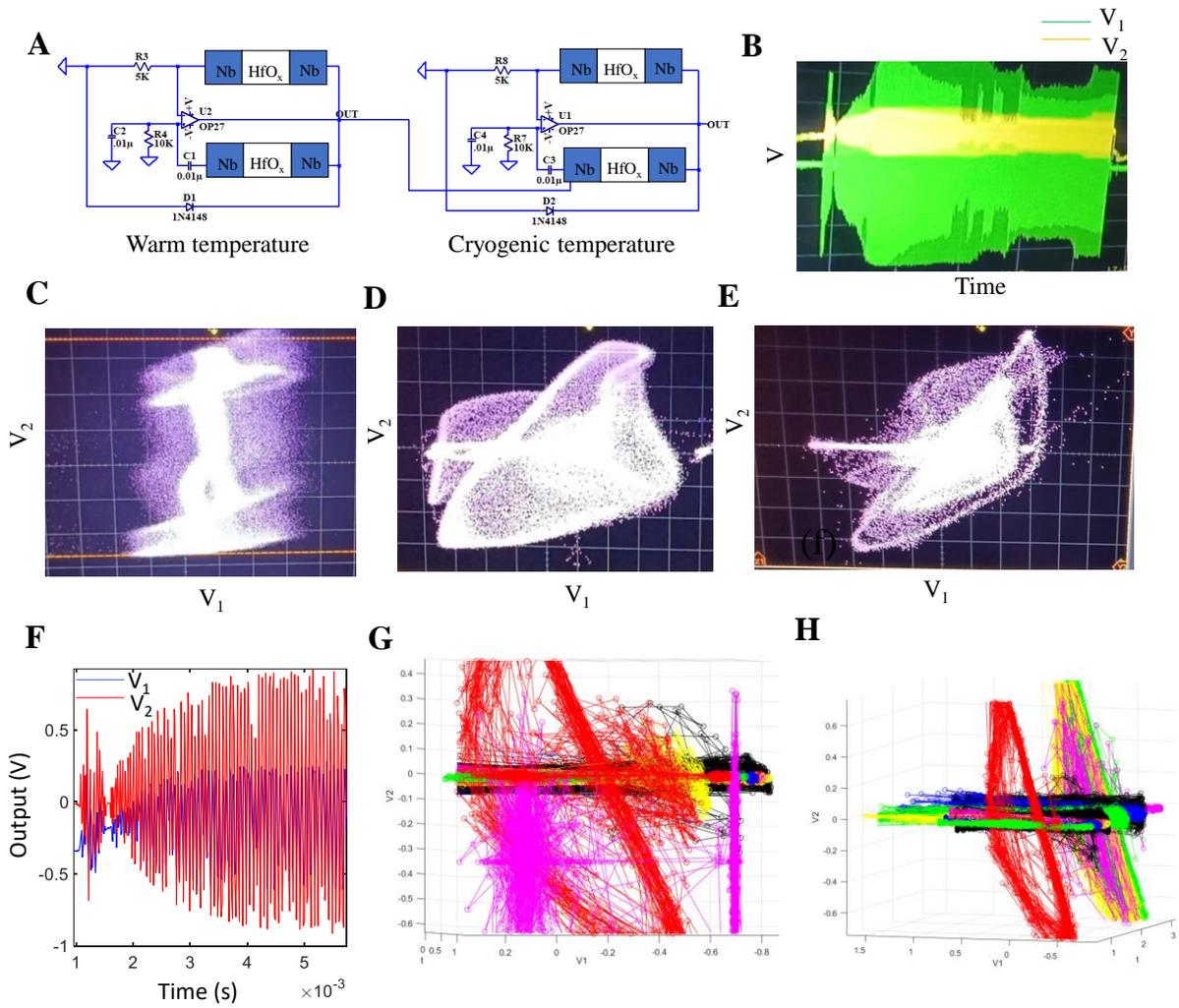

**Fig. 4. Strong Itinerancy in networks.** (**A**) Network of two artificial neurons with the memories of neuron#1 at warm/room temperatures and neuron#2 cryogenically cooled into the superconducting state at 8.1 Kelvin (**B**) Oscilloscope capture example in the time-domain (**C-E**) Examples of data collected in the phase-plane under conditions that produce strong itinerant behavior (**F**) Sample time-domain response for the combined output of the two artificial neurons while under strong itinerant conditions and (**G, H**) Examples of 3D plots in the phase-plane showing how time is spent in an attractor mode followed by entering into a new attractor mode.

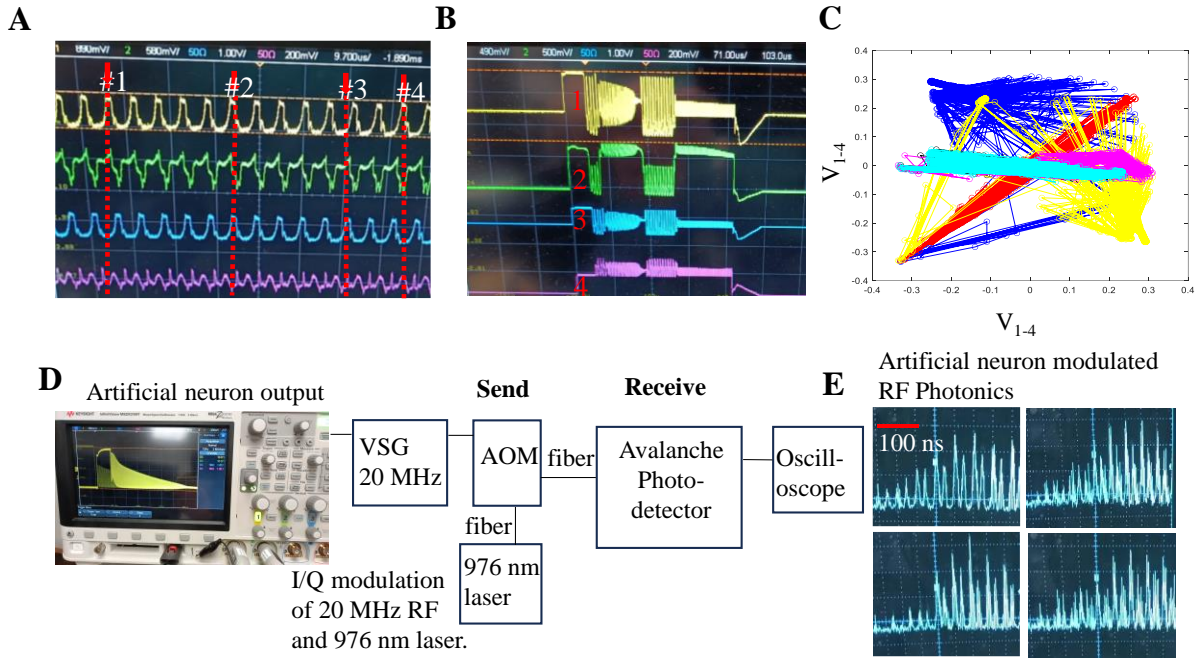

**Fig. 5. Applications.** (**A**) Oscilloscope capture in a 4-artificial neuron experimental feedback network and the time-domain output of all 4 neurons showing with arrows the changes in the level of phase-shift between the outputs of the various neurons that can be used to implement gait control or other central patterns for biologically inspired intelligent control (**B**) High resolution output in the time-domain showing the autonomous learning capability of such a network due to the adaptive and itinerant properties of the artificial neurons (**C**) Phase plane portraits for all the combinations at the output voltage nodes (**D, E**) Experimental artificial neuron modulation of RF Photonics signals, where the output of an artificial neuron is used to modulate a higher frequency RF carrier signal at 20 MHz and a 976 nm optical laser signal with an acousto-optical modulator (AOM). The modulated RF photonics signal is then detected after propagating in a fiber optic cable with an avalanche photodetector.